\documentclass[runningheads]{llncs}

\usepackage{graphicx, subcaption} \graphicspath{ {figures/} } 
\usepackage[pdftex, pdfpagelabels, hypertexnames, bookmarks, bookmarksnumbered, plainpages=false, naturalnames=false, pdfpagemode=UseOutlines]{hyperref}
\hypersetup{colorlinks, citecolor=[rgb]{0.0,0.0,0.0}, filecolor=[rgb]{0.0,0.0,0.0}, linkcolor=[rgb]{0.0,0.0,0.0}, urlcolor=[rgb]{0.0,0.0,0.0}}

\usepackage{amsmath, amssymb}
\usepackage{algorithm, algorithmic}
\usepackage{booktabs, multirow}
\usepackage{microtype}

\begin{document}

\title{Semi-Supervised Multi-Task Learning\\ With Chest X-Ray Images}

\author{Abdullah-Al-Zubaer Imran \and Demetri Terzopoulos}
\institute{Computer Science Department, University of California, Los Angeles, CA, USA
}
\authorrunning{A.-A.-Z.~Imran and D.~Terzopoulos}

\maketitle

\begin{abstract}
Discriminative models that require full supervision are inefficacious in the medical imaging domain when large labeled datasets are unavailable. By contrast, generative modeling---i.e., learning data generation and classification---facilitates semi-supervised training with limited labeled data. Moreover, generative modeling can be advantageous in accomplishing multiple objectives for better generalization. We propose a novel multi-task learning model for jointly learning a classifier and a segmentor, from chest X-ray images, through semi-supervised learning. In addition, we propose a new loss function that combines absolute KL divergence with Tversky loss (KLTV) to yield faster convergence and better segmentation performance. Based on our experimental results using a novel segmentation model, an Adversarial Pyramid Progressive Attention U-Net (APPAU-Net), we hypothesize that KLTV can be more effective for generalizing multi-tasking models while being competitive in segmentation-only tasks.
\end{abstract}


\section{Introduction}

The effective supervised training of deep neural networks normally requires large pools of labeled data. In medical imaging, however, datasets tend to be limited in size due to privacy issues, and labeled data is scarce since manual annotation requires tedious, time-consuming effort by medical experts, making it not only expensive, but also susceptible to subjectivity, human error, and variance across different experts. Although some large labeled datasets are available, they can be seriously imbalanced by over-representation of common problems and under-representation of rare problems.

The success of discriminative models such as regular CNNs for classification or segmentation, depends on large labeled training datasets to make predictions about future unobserved examples. Generative modeling has recently received much attention with the advent of deep generative models, such as GANs. Since they can learn real data distributions, they are becoming increasingly popular given the abundance of unlabeled data. 

Via generative modeling, we can perform multi-task learning in a semi-supervised manner, without large labeled datasets. In practice, we train a deep learning model to perform a single task (classification, segmentation, detection, etc.) by fine-tuning parameters until its performance no longer improves. The same model can subsequently be enabled to perform better in other tasks. In fact, the domain-specific features from the related tasks are leveraged to improve the generalization of the model through multi-task learning \cite{Caruana1993MultitaskLA}. Hence, one objective regularizes another to accomplish multiple tasks within a common model. 

We introduce a novel generative modeling approach to joint segmentation and classification from limited labeled data, in a semi-supervised manner, and apply it to chest X-ray imagery. Our technical contributions are twofold: (1) a novel multi-task learning model for semi-supervised classification and segmentation from small labeled medical image datasets and (2) a new loss function combining absolute KL divergence and Tversky loss (KLTV) for semantic segmentation. 

\subsection{Related Work}

 Several single-task classification and segmentation models are available in the chest X-ray literature. Based on the popular segnet architecture, Mittal {\it et al.} \cite{mittal2018lf} proposed a fully convolutional encoder-decoder with skip connections for lung segmentation in chest X-ray images. Adversarial training of an FCN followed by a CRF has been applied to non-overfitting mammogram segmentation \cite{zhu2018adversarial}. Adversarial learning has been utilized for segmentation (semantic-aware generative adversarial nets \cite{chen2018semantic}, structure correcting adversarial nets \cite{dai2018scan}, etc.) as well as in disease classification from chest X-ray images (semi-supervised domain adaptation \cite{madani2018semi}, attention-guided CNN \cite{guan2018diagnose}, semi-supervised multi-adversarial encoder \cite{imran2019multi}).
 
Unlike the above models, our model jointly performs both classification and segmentation. Several prior efforts address multi-task learning with CNNs and generative modeling. Rezaei {\it et al.} \cite{Rezaei2018MultiTaskGA} proposed a GAN model combining a set of auto-encoders with an LSTM unit and an FCN as discriminator for semantic segmentation and disease prediction.  Girard {\it et al.} \cite{girard2019joint} used a U-Net-like architecture coupled with graph propagation to jointly segment and classify retinal vessels. Mehat {\it et al.} \cite{Mehta2018YNetJS} proposed a Y-Net, with parallel discriminative and convolutional modularity, for the joint segmentation and classification of breast biopsy images. Another multi-tasking model was proposed by Yang {\it et al.} \cite{yang2017novel} for skin lesion segmentation and melanoma-seborrheic keratosis classification, using GoogleNet extended to three branches for segmentation and two classification predictions. Khosravan {\it et al.} \cite{khosravan2018ssmtl} used a semi-supervised multi-task model for the joint learning of false positive reduction and nodule segmentation from 3D CT. Ours is the first model to pursue a multi-task learning approach to the analysis of chest X-ray images.

\section{Model Description}

\begin{figure}[t]
    \centering
    \subcaptionbox{}{
    \includegraphics[width=0.2\linewidth,trim={240 190 270 180},clip]{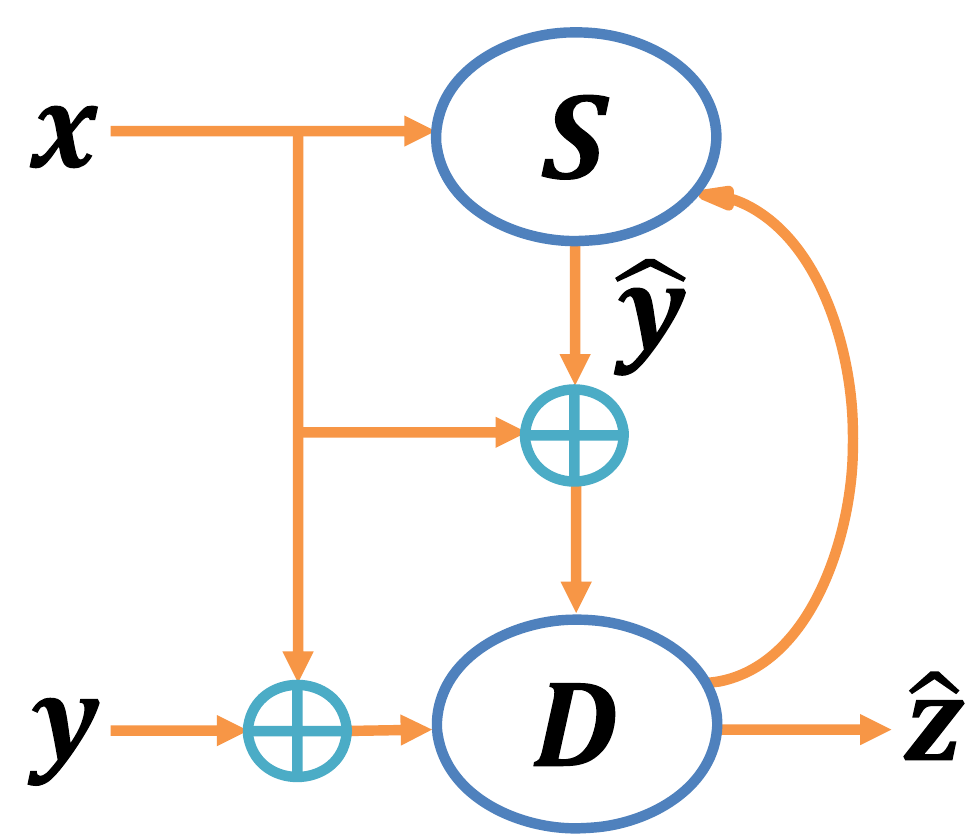}}
    \hfill
    \subcaptionbox{}{\includegraphics[width=0.74\linewidth,trim={100 210 120 190},clip]{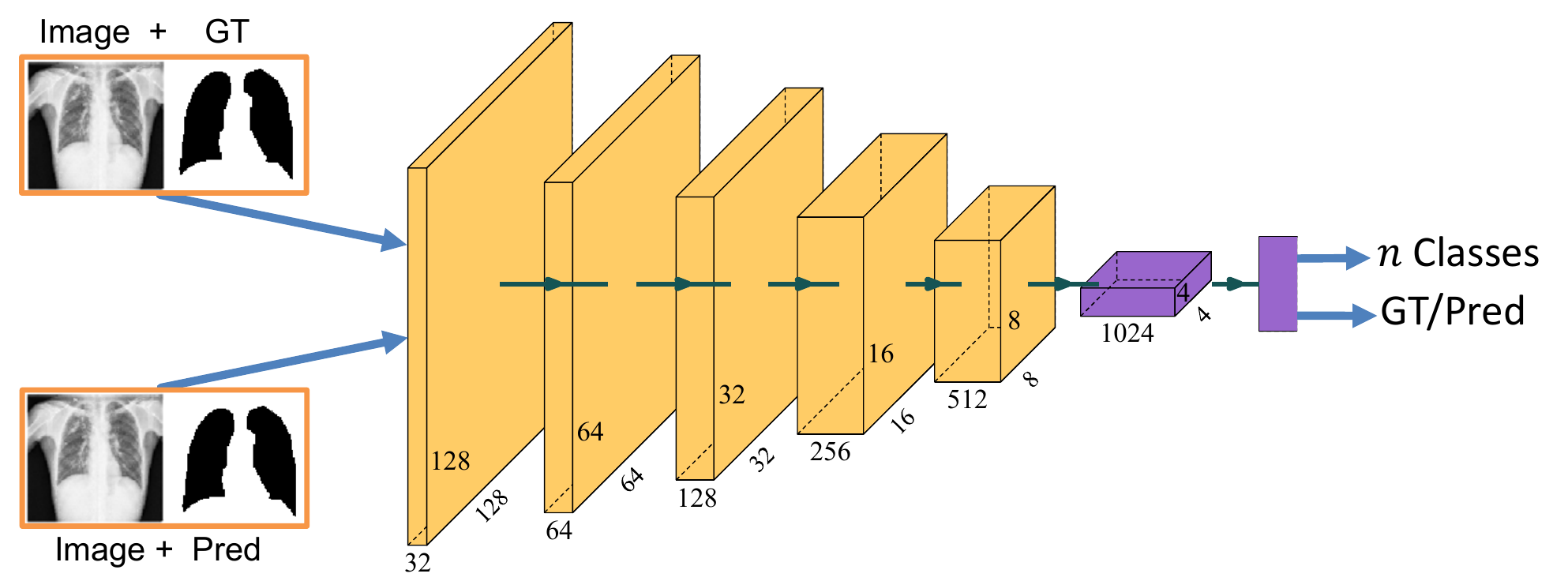}}
    \caption{(a) Basic structure of the proposed APPAU-Net model. The segmentor $S$ predicts segmentation $\hat{y}$ from a given image $x$. The discriminator $D$ predicts the class label $\hat{z}$ from image-real label pair $(x,y)$; $z = 0,\dots, n$ are real disease classes and $z=n+1$ denotes the predicted class; (b) Detailed architecture of the Discriminator $D$ (as a CNN) network of the APPAU-Net model.}
    \label{fig:appau-net}
\end{figure}

\subsection{Adversarial Pyramid Progressive Attention U-Net}

\begin{figure}[t]
    \includegraphics[width=\linewidth,trim={100 140 100 140},clip]{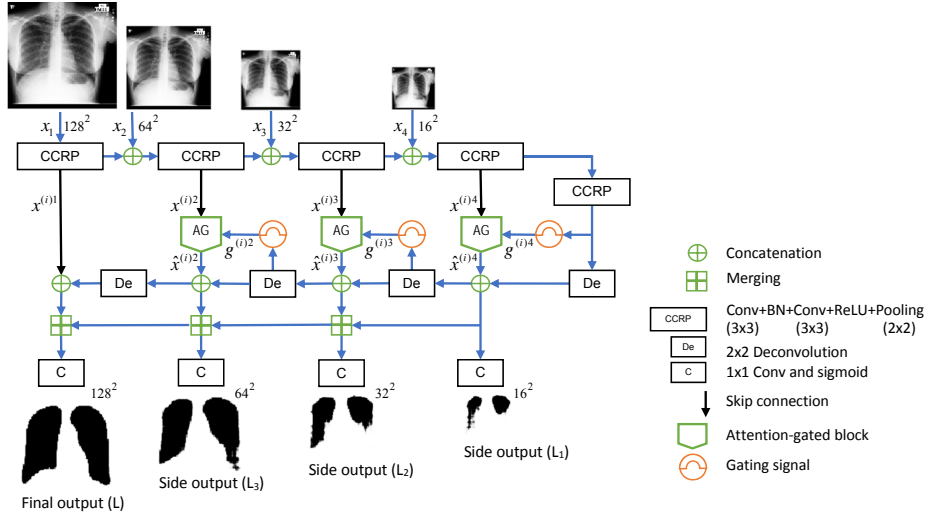}
    \caption{Architecture of the segmentor or PPAU-Net in our APPAU-Net model. The encoder takes inputs at different scales and progressively adds the side-outputs from the attention-gated decoder. The discriminator takes image-label or image-predicted label pairs and classifies the images.}
    \label{fig:ppau-net}
\end{figure}

Our proposed APPAU-Net model consists of two major building blocks, a segmentor $S$ and a discriminator $D$ (Figure~\ref{fig:appau-net}). $S$ primarily performs segmentation prediction $\hat{y}$ from a given image $x$. $S$ consists of a pyramid encoder and a progressive attention-gated decoder modifying a U-Net. The $S$ network, which is illustrated in Figure~\ref{fig:ppau-net}, receives the image input $x$ at different scales in different stages of the encoder \cite{fu2018joint}. This pyramidal input allows the model to access class details at different scales. Moreover, while lowering resolution, the model can keep track of the ROIs, avoiding the possibility of losing them after the subsequent convolutions. The pyramid input to the encoder network enables the model to learn more locally-aware features crucial to semantic segmentation.  

Following \cite{imran2018automatic}, with deep-supervision, APPAU-Net generates side-outputs at different resolutions from the decoder. The side-outputs are progressively added to the next side-outputs before reaching the final segmentation at the original image resolution. Combining pyramid inputs and progressive side-outputs helps the model perform better in segmenting small ROIs. The side-output segmentation maps $\hat{y}_i$ are compared to the ground truth mask to calculate the side-losses of varying weights (higher resolutions are usually assigned higher weights). Therefore, the final segmentation loss is calculated as
\begin{equation}
    \label{eqn:side-loss}
    L_{seg_{(x,y)}} = \sum^4_{i=1} w_iL_{(y_i, \hat{y_i})}.
\end{equation}

However, generating segmentation maps (side-outputs) at different stages of the decoder might lead to loss of spatial detail. In cases with substantial shape variability of the ROIs, this eventually incurs larger false positives. To tackle this problem, we adapt soft-attention gates that help draw relevant spatial features from the low-level feature maps of the pyramid encoder \cite{oktay2018attention}. Feature maps are then propagated to the high-level features to generate side-outputs at different stages of the decoder. Attention-gated (AG) modules produce attention coefficients $\alpha \in [0,1]$ at each pixel $i$ that scale input feature maps $x^{(i)l}$ at layer $l$ to semantically relevant features $\hat{x}^{(i)l}$. A gating signal from coarser resolution, serves to determine the focus regions through the computation of intermediate maps, as follows:
\begin{equation}
    G_\text{attn}^l = \psi^T(\sigma(w_x^Tx^{(i)l} + w_g^Tg^{(i)l} + b_g)) + b_\psi.
\end{equation}
The linear attention coefficients are computed by element-wise summation and a $1\times1$ linear transformation. 
The parameters are $w_x$, $w_g$,  $b_g$, and $b_\psi$. The intermediate maps are then transformed using ReLU $\sigma_1$ and sigmoid $\sigma_2$ activations. Finally, after element-wise multiplication of the feature map $x^{(i)l}$ (via skip) and nonlinear transformation, $\hat{x}^{(i)l}$ is generated at each decoder stage.

The attention coefficients $\alpha_i$ retain the relevant features by scaling the low level query signal $x^{(i)l}$ through an element-wise product. These pruned features are then concatenated with upsampled output maps at different stages of the decoder. A $1\times1$ convolution and sigmoid activation is applied on each output map in the decoder to generate the side-outputs at different resolutions. With deep supervision and gating from the pyramid encoder, the model becomes semantically more discriminative.

\subsection{Loss Functions}

The two building blocks of our APPAU-Net model have different objectives.

\paragraph{Segmentor Loss:} As in the semi-supervised learning-scheme, the segmentor's objective is just based on the labeled samples. We employ Tversky loss, a generalization of Dice loss that weighs false negatives higher than than false positives in order to balance precision and recall. The segmentor's objective includes a segmentation loss and an adversarial loss, where the segmentor wants the discriminator $D$ to maximize the likelihood for the predicted segmentation generated by the segmentor. We combine an absolute KL divergence with a Tversky loss, proposing the new loss function
\begin{equation}
\label{eqn:seg_loss} 
  L_S = L_{S_{{seg}_{(y, \hat{y})}}} +  cL_{S_{adv_{(x,\hat{y})}}},
 \end{equation}
where $L_{S_{{seg}_{(y, \hat{y})}}} = aL_{S_{KL}} + bL_{S_{TV}}$,
with $L_{S_{KL}} = \sum_i^{m^2}| (y_{pl}{(i)} - \hat{y}_{pl}^{(i)})\log({y_{pl}^{(i)}}/{\hat{y}_{pl}^{(i)}})|$,
and 
\begin{equation}
L_{S_{TV}} = 1 - \frac{
  \sum_i^{m^2} y_{pl}^{(i)}\hat{y}_{pl}^{(i)} + \epsilon}{
  \sum_i^{m^2}y_{pl}^{(i)}\hat{y}_{pl}^{(i)} + \alpha
    \sum_i^{m^2} y_{p\bar{l}}^{(i)}\hat{y}_{pl}^{(i)} + \beta\sum_i^{m^2} y_{pl}^{(i)}\hat{y}_{p\bar{l}^{(i)}} + \epsilon
  },
\end{equation}
where $\hat{y}_{pl}(i)$ is the prediction probability that pixel $i$ is assigned label $l$ (one of the ROI labels) and $\hat{y}_{p\bar{l}}(i)$ is the probability that the pixel $i$ is assigned the non-ROI (background) label. Similarly, $y_{pl}(i)$ and $y_{p\bar{l}}(i)$ denote the pixel-wise mapping labels in the ground-truth masks. Hyper parameters $a$, $b$, $\alpha$, and $\beta$ can be tuned to weigh the KL-divergence against the Tversky loss (first pair) and weigh FPs against FNs. Small constant $\epsilon$ avoids division by zero.  The second term in the segmentor's objective is an adversarial loss, where the segmentor wants the discriminator to maximize likelihood for the paired data $x$ and predicted segmentation $\hat{y}$. Therefore, the segmentor's adversarial loss is
\begin{equation}
\label{eqn:seg_fake}  
    L_{S_{{adv_{(x,\hat{y})}}}} = - \mathbb{E}_{x,\hat{y} \sim S}\log[1 - p(z = n+1 | (x,\hat{y})].
\end{equation} 
Since the main objective of the segmentor is to generate the segmentation map, $L_{S_{adv}}$ is usually weighed using a small number $c$. 

\paragraph{Discriminator Loss:} The discriminator is trained on multiple objectives---adversary on the segmentor's output and classification of the images into one of the real classes. Since the model is trained on both labeled and unlabeled training data, the loss function of the discriminator $D$ includes both supervised and unsupervised losses. When the model receives image-label pairs $(x,y)$, it is just the standard supervised learning loss
\begin{equation}
\label{eqn:sup__class_loss}  
\begin{aligned}
    L_{D_{sup}} &= - \mathbb{E}_{x,y,z\sim p_{data}} \log[p(z = i|x,y; i< n+1)].
    \end{aligned}
\end{equation}
When it receives unlabeled data $(x,y)$ or $(x,\hat{y})$ from two different sources, the unsupervised loss combines the original adversarial losses for image-real label and image-prediction pairs: 
\begin{equation}
\label{eqn:D_real}  
    L_{D_{label}} = - \mathbb{E}_{x,y \sim p_{data}} \log [ 1 - p(z = n+1 | x,y)]
\end{equation}
and
\begin{equation}
\label{eqn:D_fake}
L_{D_{pred}} = - \mathbb{E}_{(x,\hat{y}) \sim S} \log [p(z = n+1 | x,\hat{y})].
\end{equation}

\section{Experiments and Results}

\begin{table}[t] \setlength{\tabcolsep}{4pt}
\centering
\caption{Segmentation-only performance comparison of different models in four different data setups.}
\label{table:soa}
\resizebox{\linewidth}{!}{
\begin{tabular}{clccccccccccc}
    \toprule
           Dataset
           &
           Model
           &
           DS
           &
           JS
           &
           SSIM
           &
           F1 
           &
           HD
           &
           SN
           &
           SP
           &
           PR
           &
           RC
            \\
     \midrule
         \multirow{6}{*}{\rotatebox{45}{MCX}}
           &
           U-Net-TV 
           &
           0.991 & 0.983 & 0.950 & 0.966 & 2.968 & 0.965 & 0.989 & 0.968 & 0.965 
           \\
           &
           U-Net-KLTV 
           &
           0.990 & 0.980 & 0.947 & 0.962 & 3.009 & 0.966 & 0.985 & 0.958 & 0.966 
           \\
           &
           Attention U-Net-TV 
           &
           0.984 & 0.968 & 0.922 & 0.937 & 3.768 & 0.915 & 0.987 & 0.960 & 0.915
           \\
           &
           Attention U-Net-KLTV 
           &
           0.990 & 0.980 & 0.941 & 0.960 & 3.063 & 0.957 & 0.987 & 0.962 & 0.957
           \\
           &
           PPAU-Net-TV 
           &
           0.988 & 0.978 & 0.966 & 0.958 & 3.143 & 0.967 & 0.982 & 0.949 & 0.967
           \\
           &
           PPAU-Net-KLTV 
           &
           {\bf0.992} & 0.983 & 0.949 & 0.989 & {\bf2.690} & 0.989 & 0.958 & 0.989 & 0.976
           \\
     \midrule
         \multirow{6}{*}{\rotatebox{45}{SCX}}
           &
           U-Net-TV 
           &
           0.964 & 0.931 & 0.860 & 0.955 & 4.181 & 0.975 & 0.799 & 0.936 & 0.975 
           \\
           &
           U-Net-KLTV 
           &
           0.960 & 0.923 & 0.850 & 0.950 & 4.023 & 0.963 & 0.803 & 0.936 & 0.963 
           \\
          &
           Attention U-Net-TV 
           &
           0.958 & 0.919 & 0.842 & 0.948 & 4.562 & 0.983 & 0.725 & 0.915 & 0.983
           \\
           &
           Attention U-Net-KLTV 
           &
           {\bf0.965} & 0.933 & 0.862 & 0.955 & {\bf3.684} & 0.946 & 0.894 & 0.964 & 0.946
           \\
           &
           PPAU-Net-TV 
           &
           0.954 & 0.913 & 0.838 & 0.944 & 4.523 & 0.983 & 0.700 & 0.908 & 0.983
           \\
           &
           PPAU-Net-KLTV 
           &
           0.964 & 0.930 & 0.858 & 0.954 & 3.855 & 0.961 & 0.836 & 0.946 & 0.961
           \\
    \midrule
         \multirow{6}{*}{\rotatebox{45}{JCX}}
           &
           U-Net-TV 
           &
           0.989 & 0.979 & 0.937 & 0.985 & 2.804 & 0.990 & 0.956 & 0.981 & 0.990 
           \\
           &
           U-Net-KLTV 
           &
           {\bf0.990} & 0.980 & 0.939 & 0.986 & {\bf2.553} & 0.980 & 0.988 & 0.995 & 0.977 
           \\
           &
           Attention U-Net-TV 
           &
           0.988 & 0.977 & 0.929 & 0.983 & 2.882 & 0.993 & 0.940 & 0.974 & 0.993
           \\
           &
           Attention U-Net-KLTV 
           &
           0.989 & 0.977 & 0.932 & 0.984 & 2.781 & 0.981 & 0.970 & 0.986 & 0.981
           \\
           &
           PPAU-Net-TV 
           &
           {\bf0.990} & 0.981 & 0.941 & 0.987 & 2.768 & 0.992 & 0.958 & 0.981 & 0.992
           \\
           &
           PPAU-Net-KLTV 
           &
           {\bf0.990} & 0.979 & 0.937 & 0.985 & 2.751 & 0.987 & 0.959 & 0.982 & 0.987
           \\
    \midrule  
         \multirow{6}{*}{\rotatebox{45}{CCX}}
           &
           U-Net-TV 
           &
           {\bf0.978} & 0.958 & 0.907 & 0.968 & {\bf3.322} & 0.974 & 0.928 & 0.962 & 0.974 
           \\
           &
           U-Net-KLTV 
           &
           0.969 &  0.939 & 0.874 & 0.953 & 3.502 & 0.946 & 0.926 & 0.960 & 0.946
           \\
           &
           AttnU-Net-TV 
           &
           0.970 & 0.941 & 0.878 & 0.956 & 3.643 & 0.972 & 0.883 & 0.940 & 0.972
           \\
           &
           AttnU-Net-KLTV 
           &
           0.971 & 0.943 & 0.877 & 0.956 & 3.481 & 0.944 & 0.941 & 0.968 & 0.944
           \\
           &
           PPAU-Net-TV 
           &
           0.969 & 0.940 & 0.875 & 0.955 & 3.807 & 0.978 & 0.870 & 0.934 & 0.978
           \\
           &
           PPAU-Net-KLTV 
           &
           0.967 & 0.936 & 0.868 & 0.951 & 3.472 & 0.939 & 0.932 & 0.963 & 0.939
           \\
    \bottomrule
        \end{tabular}
     }
\end{table}
\setlength{\tabcolsep}{1.4pt}

\paragraph{Dataset and Implementation Details:} 
For the supervised segmentation, we used our PPAU-Net model and KLTV as the loss function. We compared against all the preliminary segmentation models and TV loss. Then we performed semi-supervised multi-tasking for semi-supervised disease classification and lung segmentation from chest X-ray images. We used three chest X-ray datasets: the Montgomery County chest X-ray set (MCX) comprising 138 images, the Shenzhen chest X-ray set (SCX) comprising 527 images \cite{jaeger2014two}, and the JSRT dataset (JCX) comprising 247 images \cite{shiraishi2000development}. In addition, we created another dataset (CCX) comprising 912 images, by combining prior datasets. Each dataset was split into train and test sets in a 75:25 ratio and 10\% of the train set was used for model selection. Except for CCX, all the datasets were used for binary classification (normal/abnormal), while CCX was used for 3-class classification (normal, nodule, tuberculosis). The X-ray images were normalized and resized to $128\times128$ pixels. For multi-tasking, we used the Adam optimizer with momentum 0.9 and learning rates $1.0^{-5}$ $(S)$ and $1.0^{-4}$ $(D)$. Each model was trained using a mini-batch size of 16. All the convolutional layers were followed by batch-normalization, except for the convolutions that generate side-outputs. We performed dropout at a rate of 0.4 in the discriminator. Each model was evaluated after training for 300 epochs. For the classification, along with the overall accuracy, we reported the class-wise F1 scores. For the segmentation, we used the following performance metrics: Dice similarity (DS), Average Hausdorff distance (HD), Jaccard index (JI), Sensitivity (SN), Specificity (SP), F1 score, Structural Similarity Measure (SSIM), Precision (PR), and Recall (RE) scores. 

\begin{figure}[t] \def\x{0.2\linewidth}
    \centering
    \subcaptionbox{\tiny X-Ray} {\includegraphics[width=\x,height=\x]{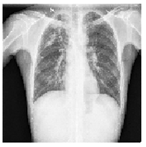}} \hfill
    \subcaptionbox{\tiny Ground Truth} {\includegraphics[width=\x,height=\x]{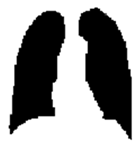}} \hfill
    \subcaptionbox{\tiny APPAU-Net-KLTV} {\includegraphics[width=\x,height=\x]{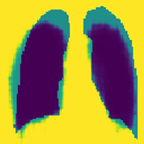}} \hfill
    \subcaptionbox{\tiny APPAU-Net-TV} {\includegraphics[width=\x,height=\x]{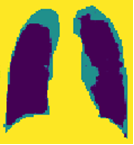}} \\[-5pt]
    \caption{Visual comparison of the lung segmentation by the APPAU-Net model with TV loss (d) and KLTV loss (c). The predicted lung mask with TV and KLTV losses are overlaid with the ground truth mask.}
    \label{fig:seg-vis}
\vspace{-1ex}
\end{figure}

\begin{table}[t] \setlength{\tabcolsep}{4pt}
\centering
\vspace{-5pt}
\caption{Performance evaluation of the APPAU-Net model for semi-supervised multi-tasking in different data settings.}
\label{table:class-seg}
\resizebox{\linewidth}{!}{
\begin{tabular}{c l c c c | c c c c c c c c c c}
            \toprule
           \multirow{2}{*}{Dataset}
           &
           \multirow{2}{*}{Model}
           &
             \multicolumn{3}{c|}{Classification}
           &
           \multicolumn{9}{c}{Segmentation}\\
           
           &
           &
           Acc
           &
           PR
           &
           RE
           &
           DS
           &
           JI
           &
           SSIM
           &
           F1 
           &
           HD
           &
           SN
           &
           SP
           &
           PR
           &
           RE
            \\
           \midrule
           \multirow{3}{*}{\rotatebox{45}{MCX}}
           &
           APPAU-Net-TV 
           &
           {\bf0.571} & 0.690 & 0.290 & 0.956 & 0.916 & 0.815 & 0.814 & 4.514 & 0.800 & 0.988 & 0.953 & 0.800 
           \\
           &
           APPAU-Net-XETV
           &
           0.514 &  0.620 & 0.280 & 0.929 & 0.868 & 0.788 & 0.778 & 4.554 & 0.903 & 0.856 & 0.684 & 0.903
           \\
           &
           APPAU-Net-KLTV 
           &
           0.543 & 0.680 & 0.200 & {\bf0.974} & 0.950 & 0.880 & 0.898 & {\bf3.914} & 0.857 & 0.944 & 0.944 & 0.857 
           \\
           \hline
           \multirow{3}{*}{\rotatebox{45}{JCX}}
           &
           APPAU-Net-TV 
           &
           {\bf0.758} & 0.000 & 0.860 & 0.972 & 0.945 & 0.864 & 0.963 & 3.755 & 0.996 & 0.831 & 0.929 & 0.996 
           \\
           &
           APPAU-Net-XETV
           &
           {\bf0.758} & 0.000 & 0.860 & 0.975 & 0.952 & 0.878 & 0.966 & 3.489 & 0.995 & 0.857 & 0.939 & 0.994
           \\
           &
           APPAU-Net-KLTV 
           &
           {\bf0.758} & 0.000 & 0.860 & {\bf0.976} & 0.953 & 0.885 & 0.966 & {\bf3.351} & 0.975 & 0.904 & 0.958 & 0.975 
           \\
           \midrule
           \multirow{3}{*}{\rotatebox{45}{SCX}}
           &
           APPAU-Net-TV 
           &
           0.477 & 0.580 & 0.300 & 0.883 & 0.790 & 0.713 & 0.877 & 6.601 & 0.999 & 0.162 & 0.782 & 0.992 
           \\
           &
           APPAU-Net-XETV
           &
           {\bf0.553} & 0.670 & 0.290 & 0.889 & 0.800 & 0.720 & 0.882 & 6.372 & 0.997 & 0.205 & 0.791 & 0.997
           \\
           &
           APPAU-Net-KLTV 
           &
           0.508 & 0.530 & 0.490 & {\bf0.921} & 0.853 & 0.746 & 0.910 & {\bf4.368} & 0.992 & 0.434 & 0.841 & 0.992 
           \\
           \midrule
           \multirow{3}{*}{\rotatebox{45}{CCX}}
           &
           APPAU-Net-TV 
           &
           {\bf 0.776} & 0.800&0.780&0.874 & 0.777 & 0.682 & 0.845 & 5.375 & 0.936 & 0.576 & 0.770 & 0.959 
           \\
           &
           APPAU-Net-XETV
           & 
           0.732 & 0.81 & 0.70 & 0.923 & 0.862 & 0.768 & 0.890 & 4.692 & 0.974 & 0.632 & 0.823 & 0.954 
           \\
           &
           APPAU-Net-KLTV 
           &
           0.750 & 0.770 & 0.750 & {\bf0.926} & 0.863 & 0.780 & 0.903 & {\bf 4.669} & 0.979 & 0.645 & 0.838 & 0.953 
           \\
           \bottomrule
        \end{tabular}
}
\end{table}

\vspace{-5pt}

\paragraph{Segmentation-Only:} At first, we evaluated the performance of our PPAU-Net model for the segmentation-only task, and compared with the baseline models incrementally. Table~\ref{table:soa} reports the performance measures of different models with varying choices of loss (TV and KLTV), showing that our model is competitive.

\vspace{-5pt}

\paragraph{Semi-Supervised Multi-Task Learning:} In the semi-supervised setting, we applied our new APPAU-Net model. Along with TV loss, we used cross-entropy with TV (XETV) loss and the proposed KLTV loss. 10\% labeled and 90\% unlabeled training data were used for every dataset. Table~\ref{table:class-seg} shows that for all four datasets the APPAU-Net model with the new KLTV loss consistently outperformed the APPAU-Net model with TV and XETV losses in both overlap and distance measures, and suggests that the model with KLTV loss generalizes better in multi-task learning. While both TV and XETV losses tend to lose some accuracy because of the additional classification task, KLTV still achieves good accuracy, comparable to fully-supervised segmentation models in Table~\ref{table:soa} and LF-segnet \cite{mittal2018lf}. Figure~\ref{fig:seg-vis} shows the segmented lungs by different models, confirming the superior performance of our APPAU-Net with KLTV loss compared to the TV loss.

\section{Conclusions}

Generative modeling provides unique advantages for learning from small labeled datasets. With adversarial training, we can perform multi-task learning to concurrently accomplish multiple objectives. We proposed and demonstrated in different settings the performance of a novel semi-supervised multi-task learning model for joint classification and segmentation from a limited number of labeled chest X-ray images. Our experimental results confirm that our APPAU-Net model even against the single-task learning of fully supervised models.

\bibliography{reference}
\bibliographystyle{splncs}

\appendix

\section{Data Description} For performance evaluation of the proposed model and a list of other models as baselines, we made use of the following three publicly available datasets: the Montgomery County chest X-ray set (MCX), the Shenzhen chest X-ray set (SCX) available from NIH \cite{jaeger2014two}, and the JSRT dataset available from Japanese Society of Radiological Technology \cite{shiraishi2000development} (JCX). An additional dataset was created combining all of them (CCX). Table \ref{table: data-split} shows the partitioning of the datasets for the four different settings. We used or prepared the images as follows:
\begin{enumerate}
    \item 
    MCX: In this dataset, there are 138 frontal X-Rays: 80 X-Rays are normal and 58 X-Rays show manifestations of Tuberculosis. This dataset contains separate left and right lung ground truth masks, which we combined in our experiments.   
  
    \item
    SCX: This dataset comprises 662 frontal chest X-rays. Of them, 336 are normal X-Rays and 326 are abnormal cases with manifestations of TB, including pediatric X-rays. After carefully examining all the cases, we selected 527 X-rays in good agreement with the corresponding ground truth lung masks: 248 normal and 279 x-rays with abnormalities. 
    
    \item
    JSRT:  This dataset contains 247 chest X-rays in which 154 images show pulmonary lung nodules and 93 images show no lung nodules. In addition to the lung masks (separated left-right), this dataset includes ground truth masks for the heart and clavicles (separated left-right).
   
    \item
    CCX: Combining the above three datasets, we created a dataset of 912 X-ray images, that we dubbed the CCX (Combined chest X-ray) dataset. We split it into three sets: training set (615), validation set (69), and testing set (228). The models were trained on the training set and the validation set was used to determine the hyperparameters and model selection. The models were evaluated on the test set. In the combined dataset, we performed 3-class classification: normal, abnormal with TB, and abnormal with lung nodule.  
\end{enumerate}

\begin{table}[t]
    \centering
     \caption{Partitioning of the image datasets.}
     \label{table: data-split}
    \begin{tabular}{c|c|c|c|c}
        \cline{1-5}
         Dataset & Train & Validation & Test & Classes \\
         \hline
         MCX (138) & 93 & 10 & 35 & normal, TB
         \\
         SCX (527) & 355 & 40 & 132 & normal, TB
         \\
         JCX (247) & 166 & 19 & 62 & normal, Nodule
         \\
         CCX (912) & 615 & 69 & 228 & normal, TB, Nodule
         \\
         \hline
    \end{tabular}
\end{table}

\begin{algorithm}[t]
\caption{Adversarial Pyramid Progressive Attention U-Net (APPAU-Net) Training Procedure.
$m$ is the number of samples and $b$ is the minibatch-size.}
\label{alg:appau-net}
\begin{algorithmic}
\STATE $steps \leftarrow \frac{m}{b}$
\FOR{each {\bf epoch}}
\FOR{each step in $steps$}
\STATE Sample minibatch $y_i;{y^{(1)},\dots,y^{(m)}}, y_i\sim p_\text{data}(y)$;
\STATE Sample minibatch $x_i; {x^{(1)},\dots,x^{(m)}}, x_i\sim p_\text{data}(x)$;
\STATE Update discriminator $D$ by ascending along its gradient:
\begin{equation*} \nabla_{\theta_{D}} \frac{1}{m}\sum_{i=1}^m\left[\log D(x_i, y_i) + \log(1 - D(x_i, S(x_i))\right];
\end{equation*}
\STATE Sample minibatch $x_i; {x^{(1)},\dots,x^{(m)}}, x_i\sim p_\text{data}(x)$;
\STATE Update the segmentor S by descending along its gradient from the discriminator $D$ and the segmentation loss (depending on the choice of loss function):
\begin{equation*} \nabla_{\theta_{S}} \frac{1}{m}\sum_{i=1}^m \left[\log(1 - D(x_i, S(x_i))) + L_{{seg}_{(i)}}\right];
\end{equation*}
\ENDFOR
\ENDFOR
\end{algorithmic}
\end{algorithm}

\section{Semi-Supervised Learning}

The overall training procedure for our APPAU-Net model is presented in Algorithm~\ref{alg:appau-net}. The real samples and labels to $S$ are presented in the forward pass. In the backward pass, the feedback from $D$ (Figure \ref{fig:appau-net}) is passed to $S$. In the original image generator GAN, $D$ works as a binary classifier---it classifies the input image as real or synthetic. In order to facilitate the training for a $n$-class classifier, the role of $D$ is changed to an $(n+1)$-classifier. For multiple logit generation, the sigmoid function is replaced by a softmax function. Now, it can receive image-label $(x, y)$ and image-prediction $(x, \hat{y})$ pairs as inputs, and it outputs an $(n+1)$-dimensional vector of logits $\{{l}_1, {l}_2,\dots,{l}_{n+1}\}$. These logits are finally transformed into class probabilities for the final classification. Class ${(n+1)}$ is for the image-prediction pairs and the remaining $n$ are for the multiple labels in the real image-label pairs. The probability of the $(x, y)$ pair being predicted is
\begin{equation}
\label{eqn:fake_prob}  
    p(z = n+1 | (x, y)) = \frac{\exp(l_{n+1})}{\sum_{j=1}^{n+1}\exp(l_j)},
\end{equation}
and the probability that the $(x, y)$ pair is real and belongs to class $i$ is
\begin{equation}
\label{eqn:real_prob}  
    p(z= i|(x,y), i< n+1) = \frac{\exp(l_i)}{\sum_{j=1}^{n+1}\exp(l_j)}.
\end{equation}

As a semi-supervised classifier (D) and segmentor (S), the model takes labels only for a small portion of training data. For the labeled data pairs, it is like supervised learning, whereas it learns in an unsupervised manner for the unlabeled data. The advantage comes from predicting segmentation labels by the segmentor. The model learns the classifier in an adversarial manner by generating segmentation maps for the unlabeled image-label pairs.

\begin{figure}[t]
\centering
\resizebox{\linewidth}{!}{
  \begin{tabular}{c c c c c c c c c}
    & & & &
    \includegraphics[width=.4\textwidth, trim={2cm, 3cm, 2cm, 1.5cm}, clip]{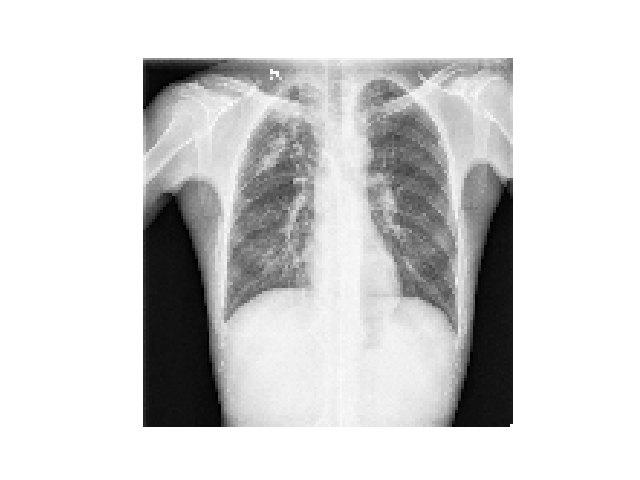} &
    \includegraphics[width=.4\textwidth, trim={2cm, 3cm, 2cm, 1.5cm}, clip]{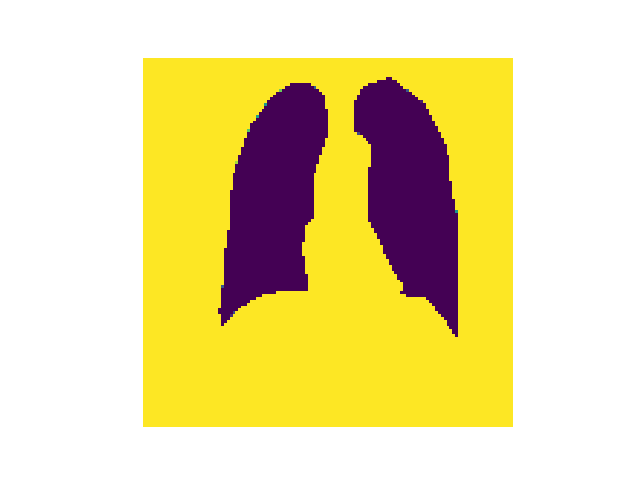} &
       \\
   & & & & \Large{X-Ray Image} & \Large{Lung Mask} &  
    \\
    \smallskip
    \\
    \rotatebox{90}{\Large{XE}} &
    \includegraphics[width=.4\textwidth, trim={2cm, 3cm, 2cm, 1.5cm}, clip]{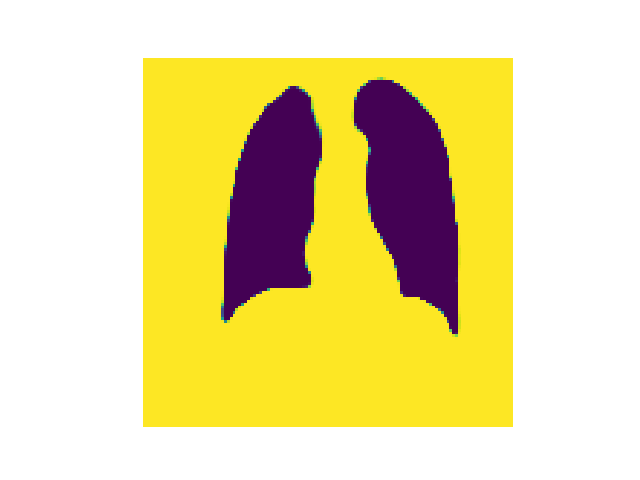} &
    \includegraphics[width=.4\textwidth, trim={2cm, 3cm, 2cm, 1.5cm}, clip]{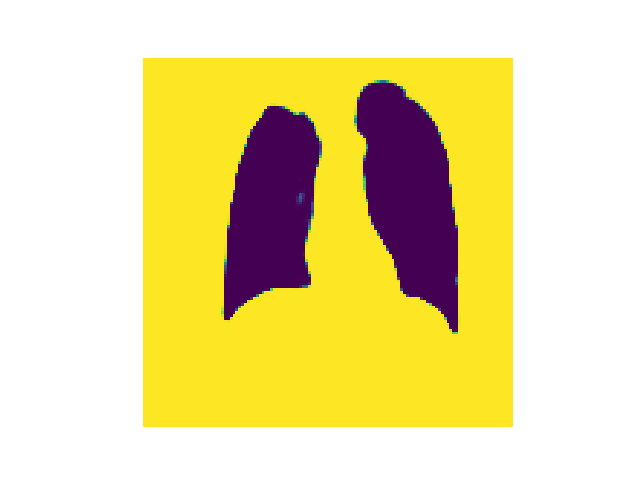} &
    \includegraphics[width=.4\textwidth, trim={2cm, 3cm, 2cm, 1.5cm}, clip]{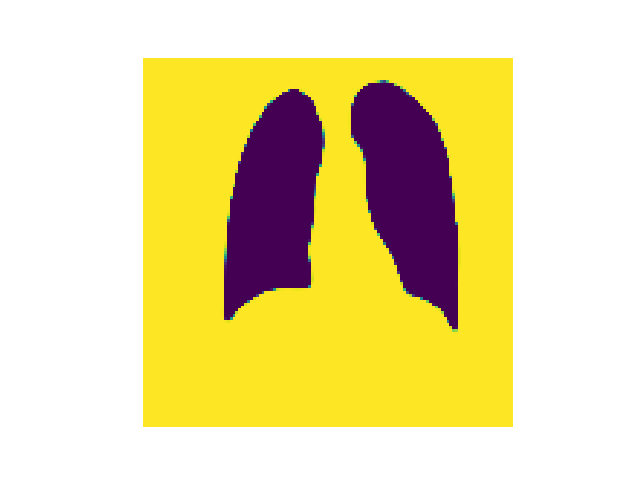} &
    \includegraphics[width=.4\textwidth, trim={2cm, 3cm, 2cm, 1.5cm}, clip]{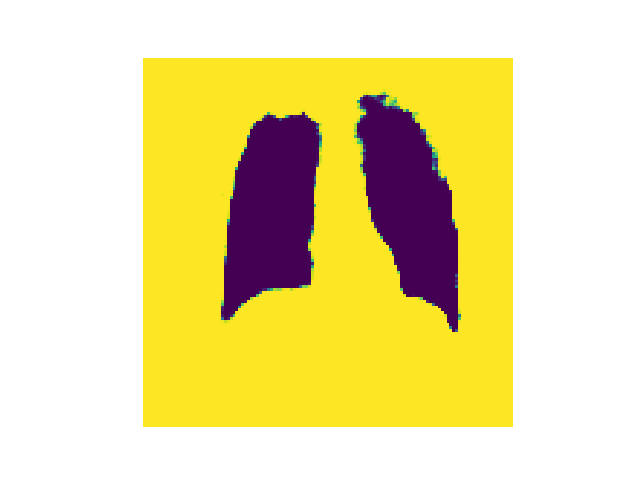}&
    \includegraphics[width=.4\textwidth, trim={2cm, 3cm, 2cm, 1.5cm}, clip]{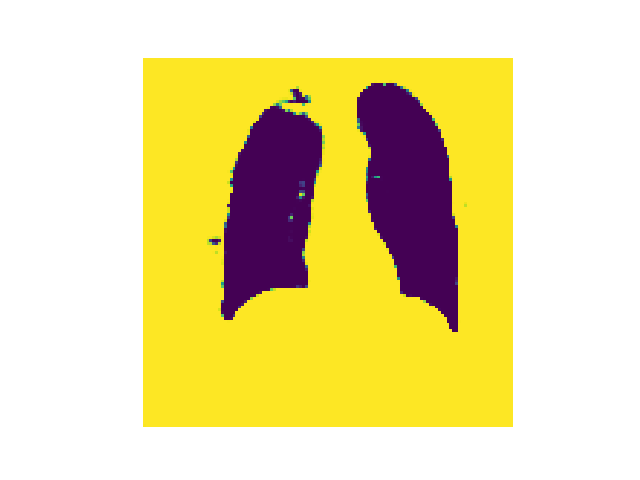}&
    \includegraphics[width=.4\textwidth, trim={2cm, 3cm, 2cm, 1.5cm}, clip]{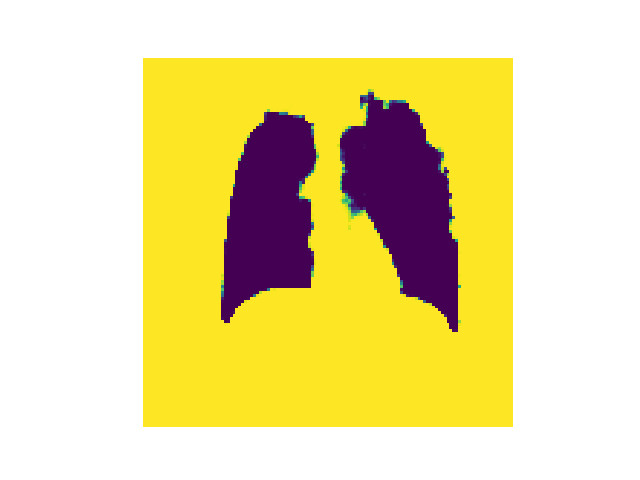}&
    \includegraphics[width=.4\textwidth, trim={2cm, 3cm, 2cm, 1.5cm}, clip]{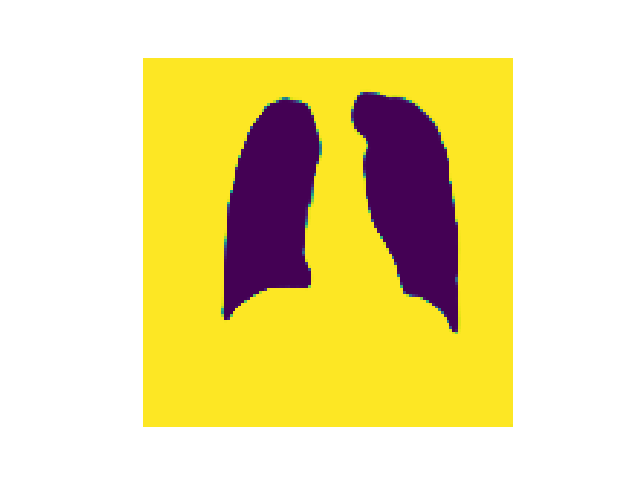}&
    \includegraphics[width=.4\textwidth, trim={2cm, 3cm, 2cm, 1.5cm}, clip]{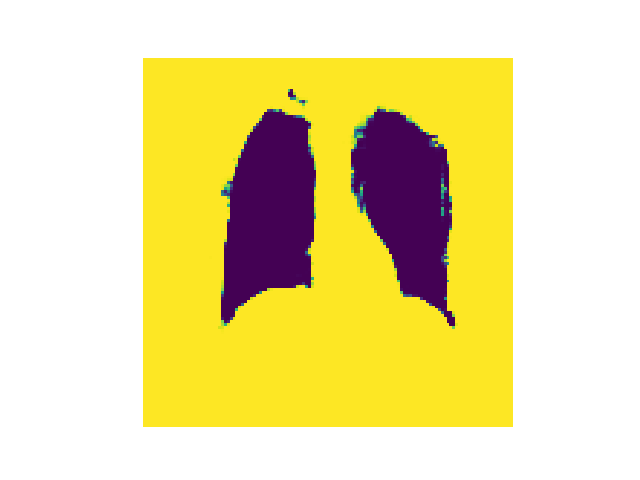}
    \\
    & \Large{U-Net} & \Large{PU-Net} & \Large{ProgU-Net} & \Large{AU-Net} & \Large{PAU-Net} & \Large{ProgAU-Net} & \Large{PPU-Net} & \Large{PPAU-Net}
    \\
    \smallskip
    \\
    \rotatebox{90}{\Large{DICE}} &
    \includegraphics[width=.4\textwidth, trim={2cm, 3cm, 2cm, 1.5cm}, clip]{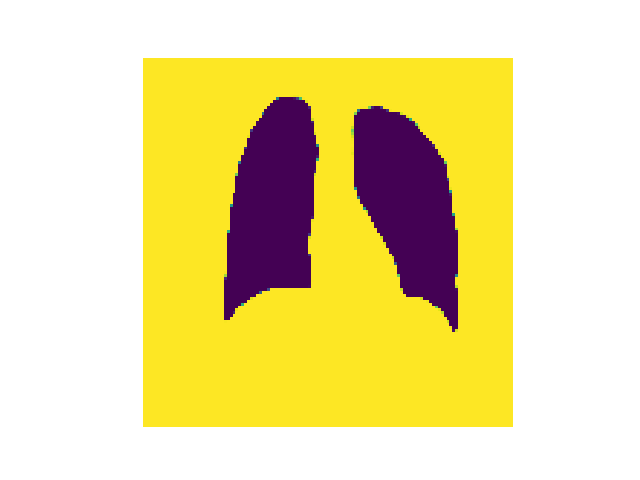} &
    \includegraphics[width=.4\textwidth, trim={2cm, 3cm, 2cm, 1.5cm}, clip]{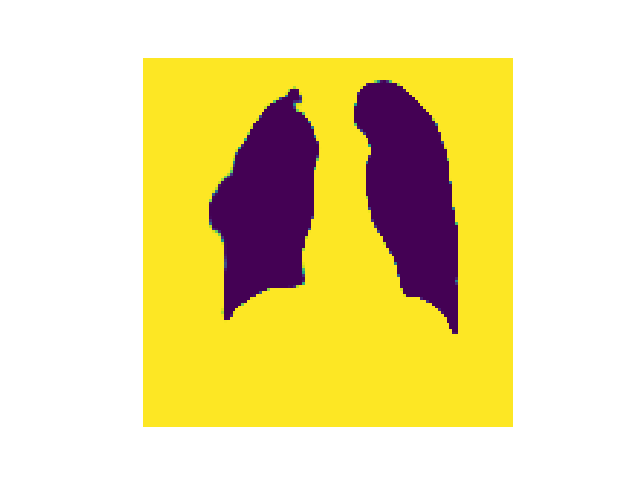} &
    \includegraphics[width=.4\textwidth, trim={2cm, 3cm, 2cm, 1.5cm}, clip]{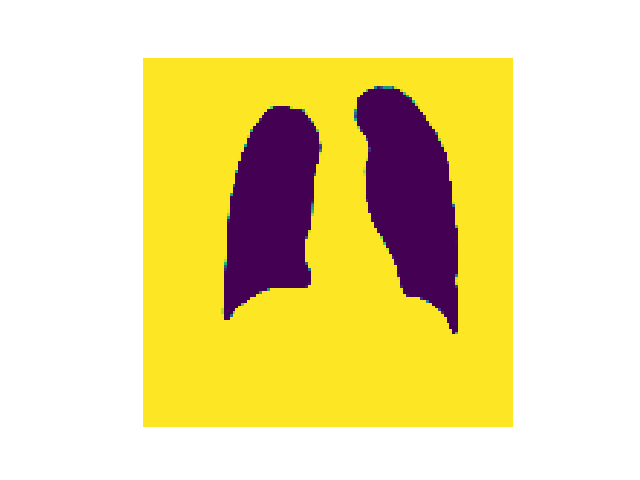} &
    \includegraphics[width=.4\textwidth, trim={2cm, 3cm, 2cm, 1.5cm}, clip]{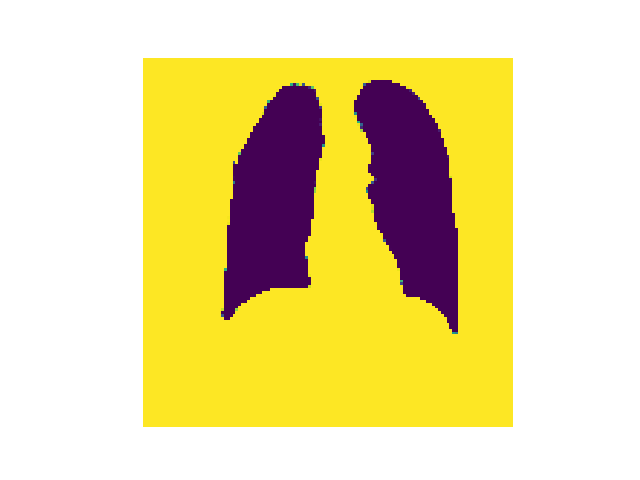}&
    \includegraphics[width=.4\textwidth, trim={2cm, 3cm, 2cm, 1.5cm}, clip]{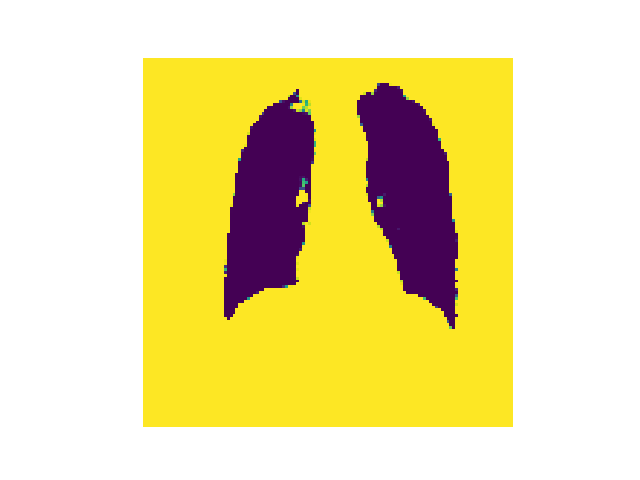}&
    \includegraphics[width=.4\textwidth, trim={2cm, 3cm, 2cm, 1.5cm}, clip]{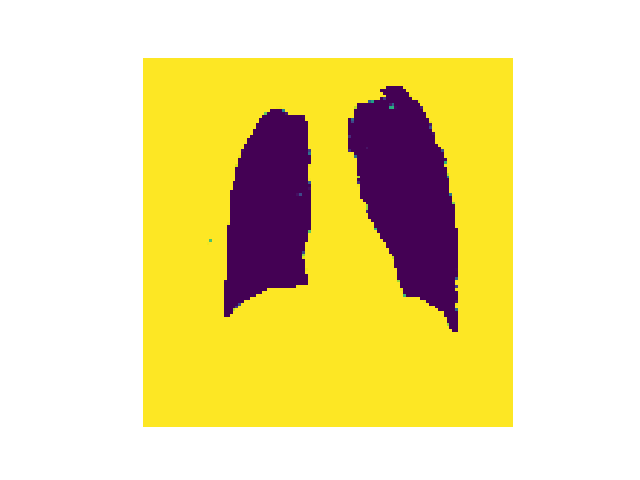}&
    \includegraphics[width=.4\textwidth, trim={2cm, 3cm, 2cm, 1.5cm}, clip]{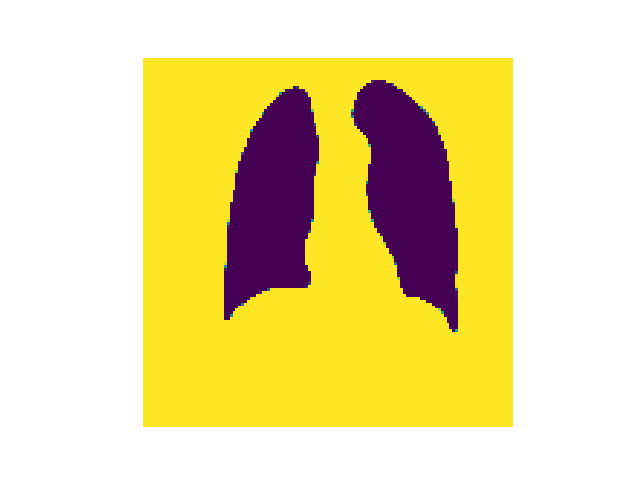}&
    \includegraphics[width=.4\textwidth, trim={2cm, 3cm, 2cm, 1.5cm}, clip]{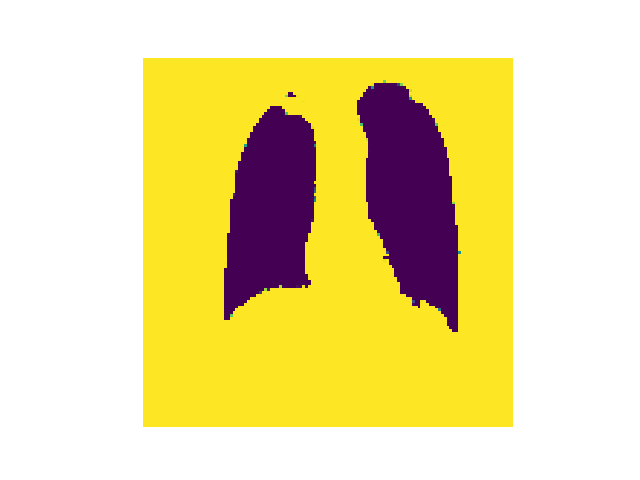}
    \\
    & \Large{U-Net} & \Large{PU-Net} & \Large{ProgU-Net} & \Large{AU-Net} & \Large{PAU-Net} & \Large{ProgAU-Net} & \Large{PPU-Net} & \Large{PPAU-Net}
    \\
    \smallskip
    \\
    
    \rotatebox{90}{\Large{TV}} &
    \includegraphics[width=.4\textwidth, trim={2cm, 3cm, 2cm, 1.5cm}, clip]{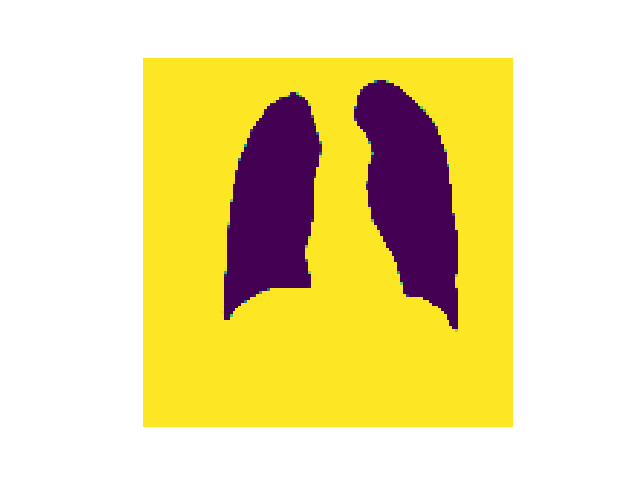} &
    \includegraphics[width=.4\textwidth, trim={2cm, 3cm, 2cm, 1.5cm}, clip]{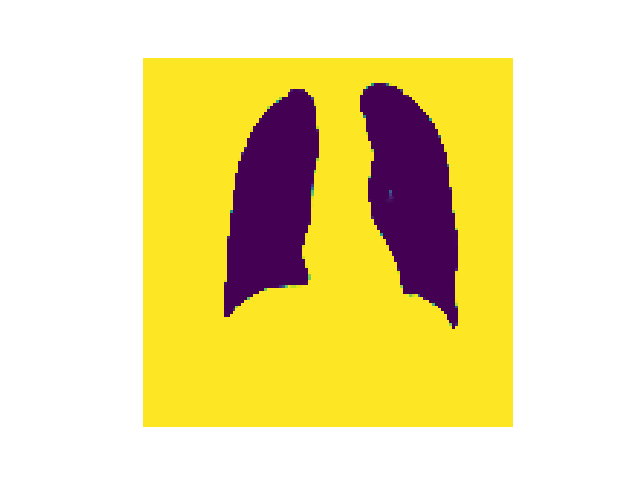} &
    \includegraphics[width=.4\textwidth, trim={2cm, 3cm, 2cm, 1.5cm}, clip]{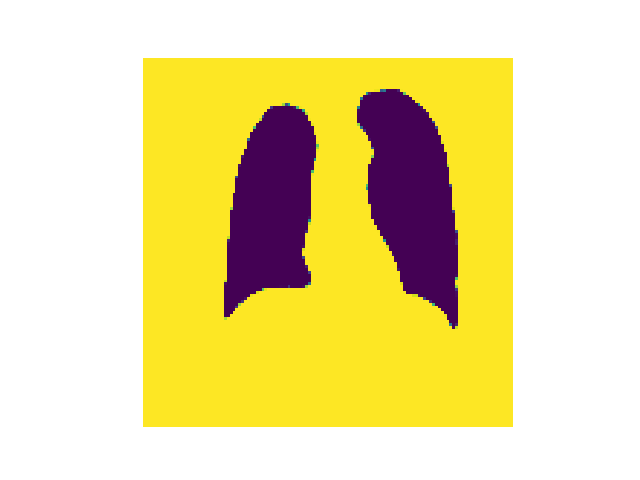} &
    \includegraphics[width=.4\textwidth, trim={2cm, 3cm, 2cm, 1.5cm}, clip]{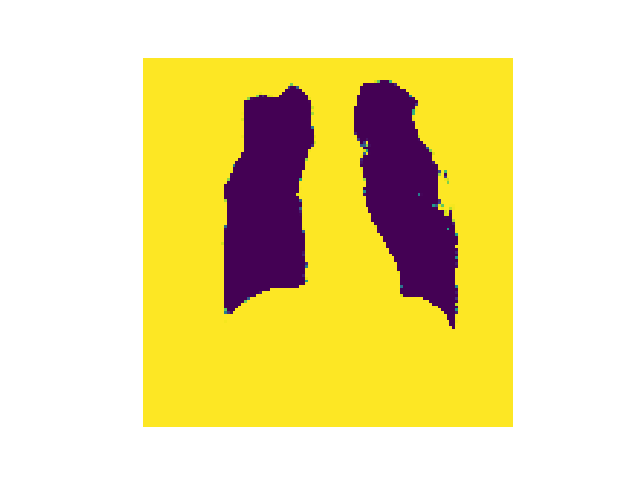}&
    \includegraphics[width=.4\textwidth, trim={2cm, 3cm, 2cm, 1.5cm}, clip]{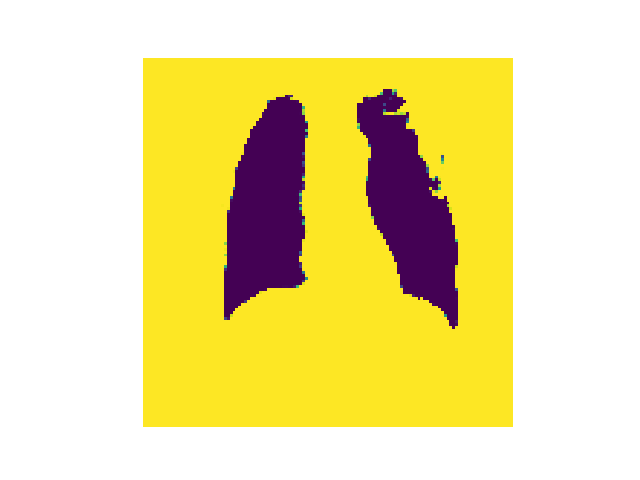}&
    \includegraphics[width=.4\textwidth, trim={2cm, 3cm, 2cm, 1.5cm}, clip]{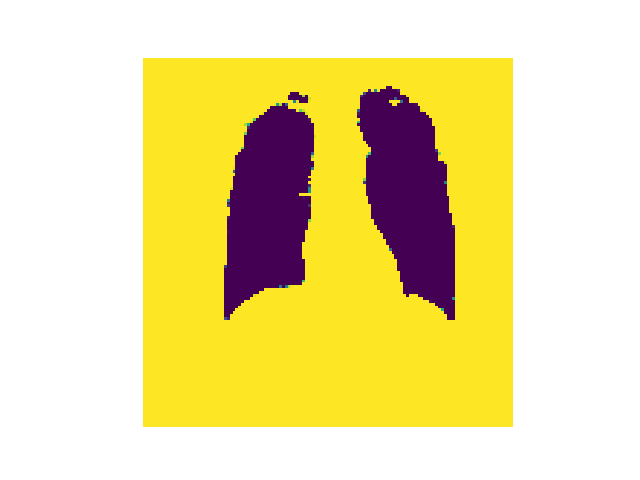}&
    \includegraphics[width=.4\textwidth, trim={2cm, 3cm, 2cm, 1.5cm}, clip]{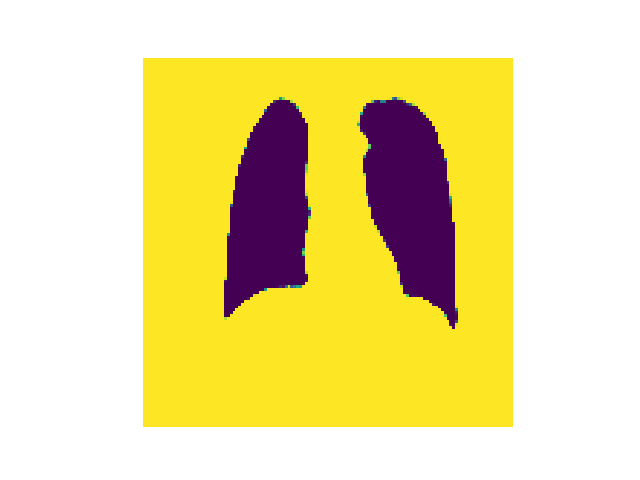}&
    \includegraphics[width=.4\textwidth, trim={2cm, 3cm, 2cm, 1.5cm}, clip]{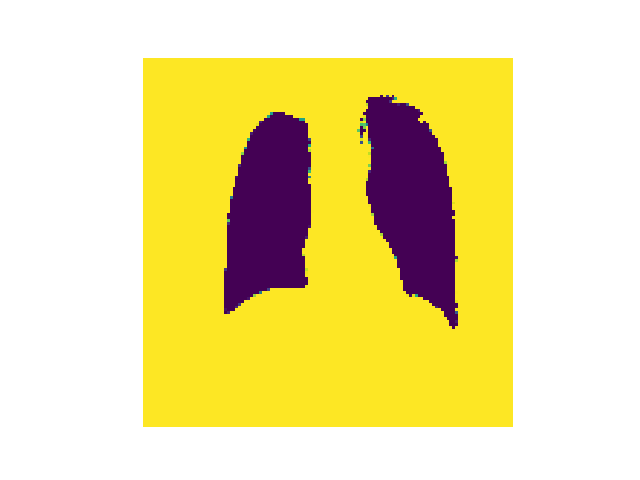}
    \\
        & \Large{U-Net} & \Large{PU-Net} & \Large{ProgU-Net} & \Large{AU-Net} & \Large{PAU-Net} & \Large{ProgAU-Net} & \Large{PPU-Net} & \Large{PPAU-Net}
    \\
    \smallskip
    \\
    \rotatebox{90}{\Large{KLTV}} &
    \includegraphics[width=.4\textwidth, trim={2cm, 3cm, 2cm, 1.5cm}, clip]{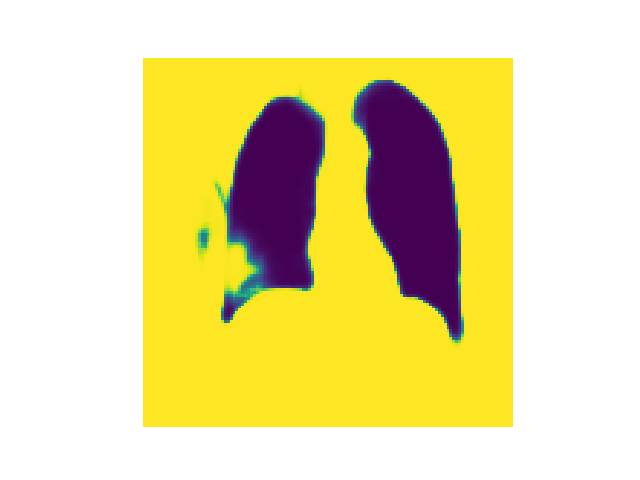} &
    \includegraphics[width=.4\textwidth, trim={2cm, 3cm, 2cm, 1.5cm}, clip]{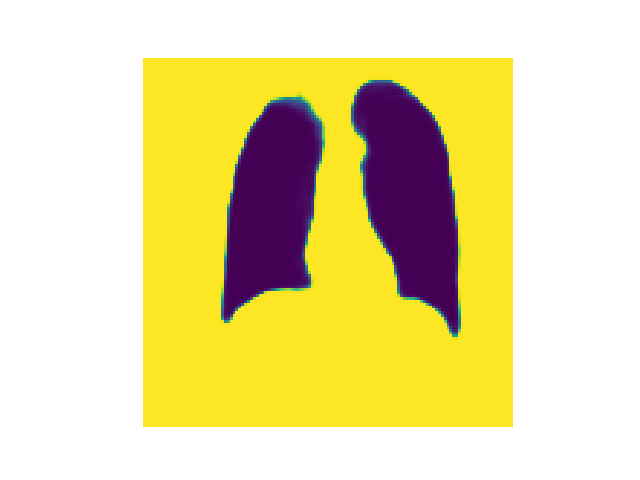} &
    \includegraphics[width=.4\textwidth, trim={2cm, 3cm, 2cm, 1.5cm}, clip]{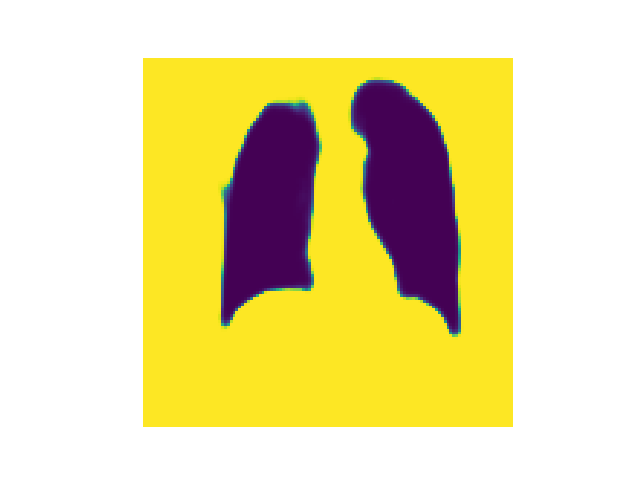} &
    \includegraphics[width=.4\textwidth, trim={2cm, 3cm, 2cm, 1.5cm}, clip]{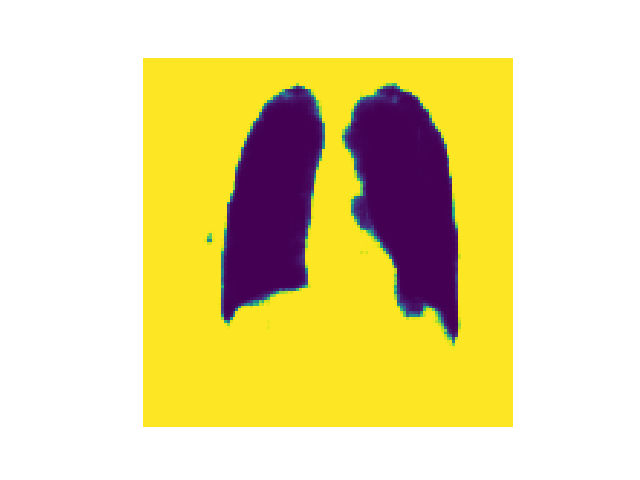}&
    \includegraphics[width=.4\textwidth, trim={2cm, 3cm, 2cm, 1.5cm}, clip]{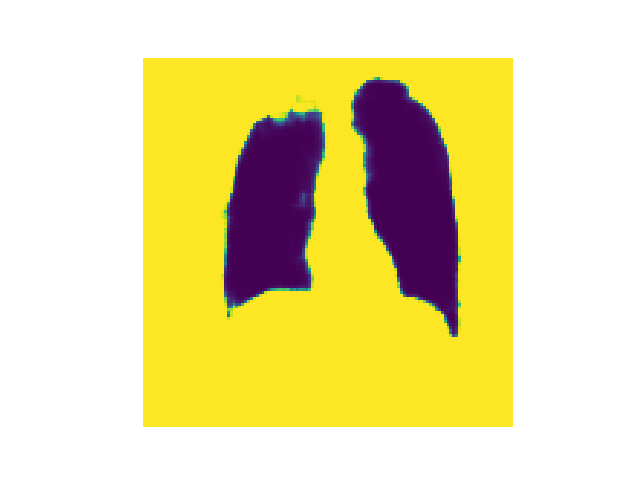}&
    \includegraphics[width=.4\textwidth, trim={2cm, 3cm, 2cm, 1.5cm}, clip]{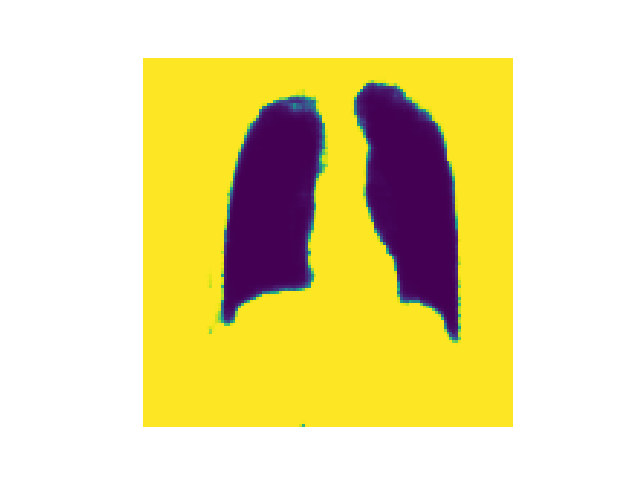}&
    \includegraphics[width=.4\textwidth, trim={2cm, 3cm, 2cm, 1.5cm}, clip]{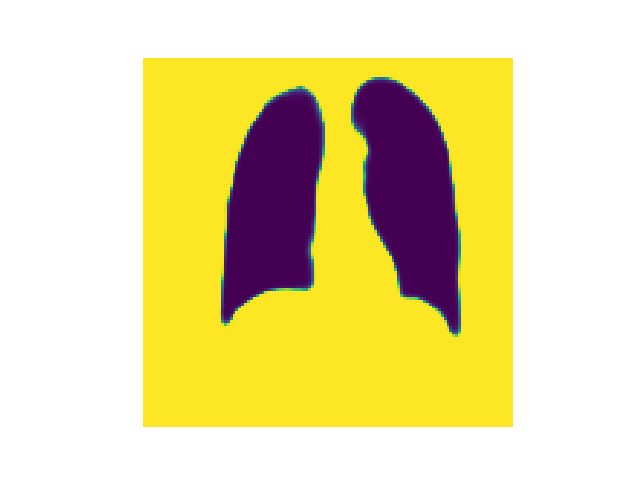}&
    \includegraphics[width=.4\textwidth, trim={2cm, 3cm, 2cm, 1.5cm}, clip]{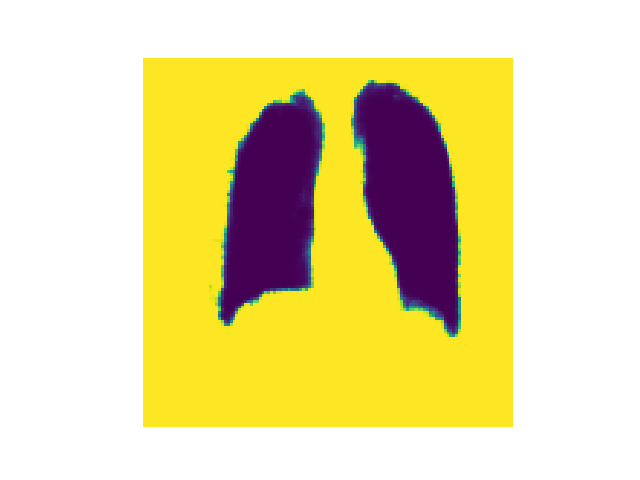}
    \\
    & \Large{U-Net} & \Large{PU-Net} & \Large{ProgU-Net} & \Large{AU-Net} & \Large{PAU-Net} & \Large{ProgAU-Net} & \Large{PPU-Net} & \Large{PPAU-Net}
    
\end{tabular}
}
  \caption{Visual comparison of the lung segmentation in an abnormal (TB) X-Ray image, for different models against the CCX dataset with varying loss functions: XE (cross-entropy loss), DICE (Dice loss), TV (Tversky loss), and KLTV (KL divergence-Tversky loss).}
  \label{fig:seg-vis1}
\end{figure}

\end{document}